\documentclass[10pt,letterpaper,copyedit]{article}
\usepackage{opex3}
\usepackage{amsmath,amsfonts,amssymb}
\usepackage{hyperref}
\hypersetup{
pdfauthor = {No\'e Curtz},
pdftitle = {Coherent frequency-down-conversion interface for quantum repeaters},
pdfsubject = {},
pdfkeywords = {},
pdfcreator = {LaTeX / hyperref},
pdfproducer = {latex}
}

\begin{document}

\title{Coherent frequency-down-conversion interface for quantum repeaters}

\newcommand{\nv}[1]{#1} 
\newcommand{\old}[1]{}  
\newcommand{\tc}[1]{#1}  

\author{No\'e Curtz$^\star$, Rob Thew, Christoph Simon$^\dagger$, Nicolas Gisin, and Hugo Zbinden}
\address{Group of Applied Physics, University of Geneva, 1211 Geneva 4, Switzerland\\$^\dagger$Institute for Quantum Information Science and Department of Physics and Astronomy, University of Calgary, Calgary T2N 1N4, Alberta, Canada}
\email{$^\star$Noe.Curtz@unige.ch}

\date{\today}

\begin{abstract*}
We report a coherence-preserving photon frequency down-conversion experiment based on difference-frequency generation in a periodically poled Lithium niobate waveguide, at the single-photon level. The coherence of the process has been demonstrated by measuring the phase coherence of pseudo single-photon time-bin qubits after frequency conversion with an interference visibility of $>$\,96\,\%. This interface could be of interest for quantum repeater based hybrid networks.
\end{abstract*}


\bibliographystyle{osajnl}
\bibliography{dfg}


\section{Introduction}

Quantum communication is the most advanced application at the single quanta level \cite{Gisin-NATP-2007}. However, while point-to-point quantum key distribution (QKD) systems become more industrialized \cite{Stucki-OE-2009}, the extension of these systems to long distances will require quantum repeater architectures \cite{Briegel-PRL-1998}. The crucial element for quantum repeaters are quantum memories (QM) \cite{Sangouard-arXiv-2009} where, currently, all but one \cite{Lauritzen-PRL-2010} operate outside of the telecommunication band. Therefore, for the majority of protocols, a coherent means of mapping these systems to fiber optic communication wavelengths is essential.

The approach chosen here is based on frequency conversion in a nonlinear crystal. Spontaneous parametric down-conversion (SPDC), sum-frequency generation (SFG) and difference-frequency generation (DFG) are all related nonlinear three-wave mixing processes. SPDC is perhaps the most common example in quantum information science and indeed provides a mechanism for coupling some absorptive quantum memories to telecom fibers \cite{Simon-PRL-2007,Usmani-NATC-2010}, although other atomic based approaches have been shown \cite{Chaneliere-PRL-2006}. However, systems that emit photons at their atomic transition do not have this option \cite{Chou-Nature-2005,Yuan-Nature-2008,Olmschenk-SCI-2009,Rosenfeld-PRL-2007}.

The process we present here, is, in a sense, the inverse of what was previously demonstrated for a SFG quantum interface (QI) \cite{Tanzilli-Nature-2005}. The importance of this change from SFG to DFG is that the SFG is used to convert the telecom source to the QM wavelength to be stored, for example by an (absorptive) QM \cite{Simon-PRL-2007,Usmani-NATC-2010}, and as such reduces the efficiency of the storage. On the other hand, DFG takes the (emissive) QM \cite{Chou-Nature-2005,Yuan-Nature-2008,Olmschenk-SCI-2009,Rosenfeld-PRL-2007} wavelength photon and converts it to the telecom band for transmission, affecting only the transmission losses. This later case has a much reduced impact on the scaling of quantum repeater architectures. Some of the leading emissive QMs use Cs (852\,nm) \cite{Chou-Nature-2005} or Rb (780\,nm) \cite{Yuan-Nature-2008} ensembles, single ions at 370\,nm \cite{Olmschenk-SCI-2009}, or atoms at 780\,nm \cite{Rosenfeld-PRL-2007}. Even at 780\,nm the losses in fiber amount to around 4\,dB\,km$^\text{-1}$ and at 370\,nm fiber transmission is simply not an option \cite{Sangouard-arXiv-2009}. As such, there is a clear need for a frequency conversion interface for a large number of the most promising atomic systems targeting quantum repeaters.

\begin{figure}[hbt]
\begin{center}
\includegraphics[width=\textwidth]{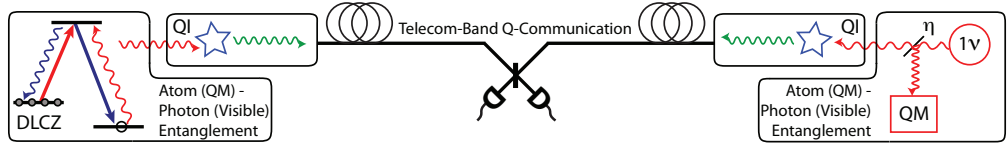}
\caption{Elementary link for entanglement creation between two quantum memories. Probability of success is enhanced as the quantum interfaces convert the photons' wavelength before propagation through the fibers. Once two adjacent elementary links are established, entanglement can be further distributed through swapping procedures.
\label{fig:qrepeater}}
\end{center}
\end{figure}

Fig.\,{\ref{fig:qrepeater}} shows a typical scenario for quantum repeaters - atom-photon entanglement is generated locally and normally some detection at a central station entangles two remote systems. Many of these elementary links can then be used to distribute entanglement over even greater distances. As we mentioned, the majority of these quantum memories operate at visible wavelengths. The frequency conversion quantum interface that we propose and demonstrate here provides a flexible means of coupling these systems to the telecom regime where transmission losses are greatly reduced. As an added advantage, systems with different wavelengths can be easily coupled by frequency-tuning the process \cite{Thew-APL-2008} and erasing the frequency information \cite{Takesue-PRL-2008} with a judicious choice of pumps and nonlinear crystals. We have pictured a DLCZ-like \cite{Duan-Nature-2001} (emissive) system on the left and a single photon source + (absorptive) quantum memory on the right \cite{Sangouard-PRA-2007} to illustrate the different types of system that could use this approach, although we note that combining such systems may also introduce temporal bandwith issues.

Optical frequency conversion has already been proposed and tested for converting telecom wavelengths to the visible regime for a wide variety of purposes. Among frequency converting interfaces, or QIs in general, teleportation \cite{Marcikic-Nature-2003} provides prohibitively low rates of successful events, limiting the conversion operating speed. SFG, or up-conversion, has been used for improved telecom single photon detection for QKD systems \cite{Thew-NJP-2006,Langrock-OL-2005,Albota-OL-2004,VanDevender-JMO-2004}. In this article we experimentally demonstrate coherent frequency conversion at the single photon level. The advantages and disadvantages of its implementation for quantum repeater schemes is discussed and we detail some of the limitations, practical and fundamental, facing this approach. 


\section{Difference-Frequency Conversion}

The DFG process provides a mechanism that takes the input waves (a photon) and mixes it with a strong pump laser, which can generate an output wave (photon) when the phase matching and energy conservation conditions are satisfied \cite{Sutherland}. We use a periodically poled Lithium niobate (PPLN) waveguide (W/G) (fabricated by the University of Nice); a potentially long interaction region and a wide variety of wavelength combinations can be realized in these devices.

In this QI experiment, we convert single-photons from 710\,nm (the signal) to 1310\,nm using a 1550\,nm pump laser. These wavelengths are chosen for expediency rather than corresponding to a particular atomic system. The frequency conversion is achieved using a nonlinear process in a type-0 PPLN W/G that is $L$\,=\,1\,cm long; has 6\,$\mu$m wide guide; has periodic poling of $\Lambda$\,=\,14\,$\mu$m; and is heated to T=\,352\,K. The W/G has previously been characterized with a normalized sum-frequency generation efficiency of around 80\,\%\,W$^\text{-1}$\,cm$^\text{-2}$ \cite{Tanzilli-Nature-2005}.

\begin{figure}[t]
\begin{center}
\includegraphics[width= .96\textwidth]{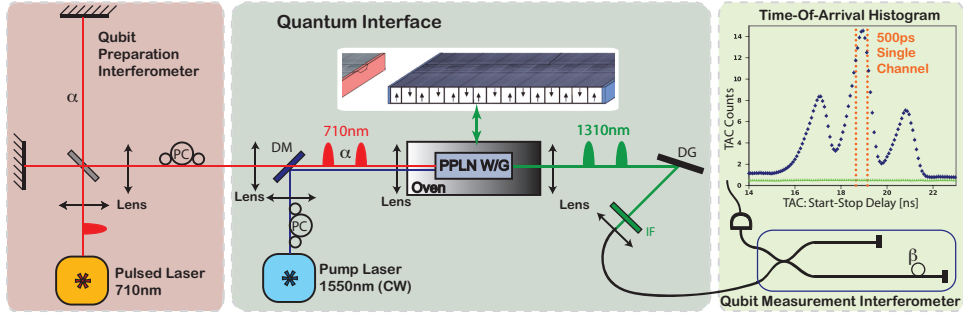}
\caption{Experimental set-up to characterize the coherence of the quantum interface (see text). A time-to-amplitude-converter (TAC) records the difference in arrival time between the pulsed laser source and the detection (inset, top right).
\label{fig:scheme}} 
\end{center}
\end{figure}

Our set-up is shown on Fig.\,{\ref{fig:scheme}}. In the first instance we characterize the efficiency of the conversion process. We bypass the two interferometers, use the 710\,nm source in the continuous-wave (cw) mode, and insert an optical attenuator before the detector. We pump the W/G with the vertically polarized 710\,nm laser diode (Toptica) in the cw mode. A tunable telecom band laser (Tunics NetTest) is amplified, filtered, vertically polarized, and combined with this on a dichroic mirror (DM) and then the wavelength is scanned to find a peak conversion efficiency ($\sim$\,1552\,nm) such that we produce quasi-phase matched photons at 1310\,nm. The photons then pass through a filtering stage before being detected by a Free-Running InGaAs/InP APD \cite{Thew-APL-2007} that has around 1.6\,kHz dark counts at 10\,\% quantum efficiency. To avoid afterpulsing we set a dead-time of 30\,$\mu$s, therefore the observed countrates have to be dead-time corrected to determine the number of 1310\,nm photons arriving on the detector.

\begin{figure*}[t]
\begin{center}
\includegraphics[width=\textwidth]{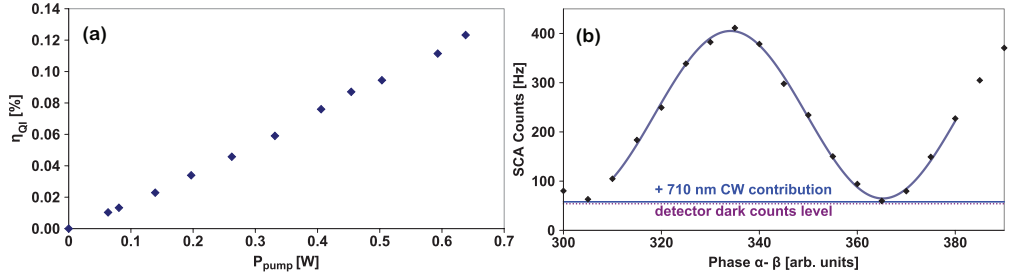}
\caption{Performance of the quantum interface. (a) QI efficiency $\eta_\text{QI}$ as a function of the 1550\,nm pump power in the waveguide. (b) Count rates for interfering events detected in the Single Channel Analyzer (SCA) of the TAC for ${<}n{>}$\,=\,1. As the phase of the fiber interferometer is scanned, a net interference visibility of V\,=\,96\,\% (raw: 84\,\%) is obtained.
\label{fig:performance}} 
\end{center}
\end{figure*}

The QI efficiency $\eta_\text{QI}$ is defined as the ratio of 1310\,nm photons in the fiber located after the filtering stage with respect to 710\,nm input photons entering the interface. The number of input photons entering the interface is determined by measuring the 710\,nm beam power (250\,$\mu$W). To determine the number of 1310\,nm photons at the output of the interface, we first filter the pump so as to ensure that the photons counted by the detector (minus dark counts) are exclusively at 1310\,nm. The filtering stage features a pump (1550\,nm) extinction $>$\,190\,dB; it consists of a diffraction grating (DG), and two interference filters (IF) centered at 1310\,nm. The beam is then coupled into a monomode fiber. This is sufficient to eliminate all residual pump photons. To avoid saturating the detection count rate we attenuate the 1310\,nm photons after conversion, with a variable fiber attenuator (Exfo). We then monitor the detector countrate; for instance with 500\,mW of pump power we set the variable attenuation to -65.3\,dB, count 6\,kHz (including 1.6\,kHz dark counts), and deduce $\eta_\text{QI}$. Fig.\,{\ref{fig:performance}a} shows $\eta_\text{QI}$ as a function of the pump power in the waveguide P$_\text{pump}$. Note that the pump laser wavelength is finely tuned as its power varies, as the beam locally heats the W/G and affects the quasi-phase matching. For maximum P$_\text{pump}$ (650\,mW) we obtain $\eta_\text{QI}$\,=\,0.13\,\%. We have evaluated the relative error on the measurements to be $\sim$\,15\,\%. We can also estimate the number of 1310\,nm photons at the output of the W/G. The DG's first-order diffraction transmits 70\,\% of the 1310\,nm light, both the IFs have a transmission of $\approx$\,80\,\%, and the coupling into the fiber has been measured at $\lesssim$\,10\,\%. Thus, about 2\,\% of the 710\,nm photons entering the W/G leave it as 1310\,nm photons.

For comparison, we determine the normalized efficiency of the W/G \cite{Sutherland}, where all the photon numbers are now considered inside the waveguide. To determine P$_\text{pump}$, we first measure the power before and after the W/G (42\,\% transmitted). Taking into account the different losses (Fresnel reflection at the W/G surfaces: 2x14\,\%, propagation in the W/G: \mbox{-0.3}\,dB\,cm$^\text{-1}$) we estimate that 60\,\% of the light is coupled into the waveguide through the input objective and can deduce P$_\text{pump}$. In the same way we calculate the signal power in the W/G and find a normalized efficiency $\eta_\text{norm}$\,$\approx$\,13\,\%\,W$^\text{-1}$\,cm$^\text{-2}$.

The principal limitation of our system is the difficulty to achieve a good mode matching, i.e. a reasonable overlap between the pump and signal input fields. This could be enhanced with the use of integrated wavelength division multiplexers and mode matching tapers \cite{Buechter-OL-2009}. Also, only $\sim$\,45\,\% of the signal beam is coupled in the W/G; with pigtailing \cite{Langrock-OL-2005,Buechter-OL-2009} and anti-reflection coatings, this could be improved to $>$\,80\,\%. Consequently, from an optimized device (W/G conversion efficiency $\sim$\,1) \cite{Chou-OL-1998}, we would only be limited by the filtering losses but could nervertheless optimistically expect an overall conversion performance $\eta_\text{QI}$\,$\gtrsim$\,50\,\% .


\section{Coherence Preservation}

To demonstrate the coherence of the DFG process, we use the complete set-up presented in Fig.\,{\ref{fig:scheme}}. The 710\,nm laser diode is pulsed at 60\,MHz, though a small cw background remains. We mention this component as it introduces a small additional noise in the final measurement. The optical pulsewidth is $\sim$\,1\,ns. We use two equally unbalanced Michelson interferometers; a bulk one before the waveguide (at 710\,nm) and a fiber one afterwards (at 1310\,nm). The path difference is $\Delta\tau$=2.2\,ns. For an initial pulse entering the first interferometer, the output can be described by the superposition (our qubit) $|\varphi \rangle= |0\rangle + e^{i\alpha}|1\rangle$, where $0$ and $1$ denote the amplitudes for the photon taking the short and the long paths in the interferometer respectively. We now attenuate the source to ensure that we have a mean photon number ${<}n{>}$\,$\approx$\,1 per qubit entering our quantum interface. The 710\,nm laser's pulses are short ($<$\,$\Delta \tau$), with a coherence that is sufficiently small so as to avoid single-photon interference phenomena. Importantly, to ensure the coherence of the overall process we need to ensure that the 1550\,nm pump coherence time is $\gg$\,$\Delta\tau$. Indeed, in the case of atomic systems, it will be necessary to ensure that the pump is more coherent than the emitted photons.

After successful frequency conversion, the qubits, now at 1310\,nm, are measured using a fiber Michelson interferometer. In the top right of Fig.\,{\ref{fig:scheme}} we see an histogram, which illustrates the different possible arrival times - the difference between the single photon detection and the oscillator driving our pulsed laser. The measurement is made with a time-to-amplitude converter (TAC). By temporally selecting the central peak (i.e. interfering short-long and long-short paths) using a single channel analyzer (SCA) we project onto the state $|\varphi' \rangle= |0\rangle + e^{i\beta}|1\rangle$. We then vary the phase in the fiber interferometer to scan the phase difference $\alpha - \beta$, producing the interference fringes that appear in Fig.\,{\ref{fig:performance}b}. We find a raw visibility of 84\,\% and a net visibility of 96\,\%. The net visibility is calculated by subtracting the detector dark counts and the contribution from the converted cw 710\,nm light, which accounts for $\sim$\,1\,\% of the peak in the single channel analyzer. The latter number has been determined by shifting the peaks out of the single channel and monitoring the contribution in the countrate on the detector due to the 710\,nm source. The widths of the peaks recorded in the histogram are around 1.3\,ns due to the 1\,ns optical pulsewidth and the 800\,ps jitter of the free-running detector (independently measured). Consequently, because of the path-length difference of the interferometers is 2.2\,ns, there is a small overlap between central and lateral peaks, slightly degrading the visibility. A small (500\,ps) SCA window is used to alleviate this problem.


\section{Discussion}

Frequency conversion experiments at the single photon level have quite frequently been plagued by excess noise photons \cite{Thew-APL-2008}. We have used a relatively simple filtering system to achieve a total extinction of the pump. A reason for the absence of any additional noise photons is that the pump wavelength is greater than the target (1310\,nm), removing the possibility of SPDC noise and reducing the chance of any Raman processes, compared to the reverse situation \cite{Thew-NJP-2006,Langrock-OL-2005}. Indeed, for any atomic system one can efficiently limit the pump noise with a suitable pump, and/or a cascaded DFG scheme \cite{Jelc-SPW-2009}. In addition, it is worth mentioning that non-converted 1310\,nm photons emitted by our pump are eliminated by a 1550\,nm IF placed after the amplifier; without this we would generate a photonic noise increasing linearly with the pump power. 

It is worthwhile to consider the potential impact of such an interface on the performance of quantum repeaters. We estimated previously that this type of interface could achieve 50\,\% overall efficiency; this amounts to 15\,km of fiber with losses of 0.2\,dB/km. The conversion loss enters linearly for the rates in repeater protocols based on single-photon entanglement (so one just loses a factor of two for 50\,\% conversion), and quadratically for protocols based on two-photon entanglement. In \cite{Sangouard-arXiv-2009}, the authors compare transmission rates for different systems, numbers of elementary links, and protocols; these all assume (implicitely) that there is a quantum interface with 100\,\% efficiency, i.e. the emitted photons are at 1550\,nm. A final remark is that although the type-0 W/G requires a well defined polarization, it doesn't prevent one implementing this scheme with polarization-encoded qubits, as solutions have previously been demonstrated with only a minor increase in complexity \cite{Martelli-OE-2009}.


\section{Conclusion}

In this paper, we have demonstrated the coherent frequency down-conversion of weak-coherent-pulse time-bin qubits. The efficiency of the current demonstration is limited to less than 1 percent; however there is great potential for gains using an interface optimized for a given atomic system (potentially $>$\,50\,\%). We find an interference visibility of $>\,96\,\%$, with no reduction due to the frequency conversion process - no contribution from the pump photons or other parasitic nonlinear processes. Currently, to the best of our knowledge, there is no other practical solution available that can perform this function with such flexibility. Indeed, this flexibility opens up the possibility of using diverse atomic systems in one quantum repeater structure to create hybrid networks that may then take advantage of different quantum technologies.


\section*{Acknowledgements}
Authors acknowledge support from the Swiss NCCR-"Quantum Photonics" and the EU FP7 project Q-ESSENCE. We thank C. Barreiro for technical support,  University of Nice for providing the waveguide and H. de Riedmatten and S. Tanzilli for useful discussions. While preparing this article similar work has been brought to our attention \cite{Takesue-PRA-2010, Ding-OL-2010}.

\end{document}